\documentclass[12pt,fleqn]{article}
\usepackage{times}
\usepackage[T1]{fontenc}
\normalfont
\usepackage{fourier} 
\usepackage{graphicx}
\usepackage{wrapfig}
\usepackage{amsmath}
\usepackage{amssymb}
\usepackage{color}
\usepackage{xcolor}
\usepackage[breaklinks=true]{hyperref}
\usepackage{ulem}
\usepackage{cite}
\newcommand{\vev}[1]{{\langle #1 \rangle}}

\newcommand{\ie}{{\it i.e.}}

\def\rew{{\rm Re}\omega}    
\def\imw{{\rm Im}\omega}

\begin{document}
\baselineskip 0.6cm
\renewcommand{\thefootnote}{\#\arabic{footnote}} 
\setcounter{footnote}{0}
%
\begin{titlepage}
\begin{center}

\begin{flushright}
KEK-TH-2287
\end{flushright}


{\Large \bf 
What if a specific neutrinoless double beta decay is absent
}\\

\vskip 1.2cm

{
Takehiko Asaka$^1$,
Hiroyuki Ishida$^2$,
and
Kazuki Tanaka$^3$
}

\vskip 0.4cm

$^1${\small 
  Department of Physics, Niigata University, Niigata 950-2181, Japan
}

$^2${\small 
	KEK Theory Center, IPNS, Tsukuba, Ibaraki 305-0801, Japan
}

$^3${\small 
Graduate School of Science and Technology, Niigata University, Niigata 950-2181, Japan}

\vskip 0.2cm

(December 24, 2020)

\vskip 2cm

\begin{abstract}
We consider the seesaw model with two right-handed neutrinos $N_1$ 
and $N_2$ which masses are
hierarchical, and investigate their contribution to the neutrinoless double beta 
($0 \nu \beta \beta$) decay.  Although the lepton number is broken by 
the Majorana masses of right-handed neutrinos, such decay processes
can be absent in some cases.  
We present a possibility where
the lighter $N_1$ gives a destructive contribution to that of active neutrinos
by choosing the specific mixing elements of $N_1$,
while $N_2$ is sufficiently heavy not to contribute to the $0 \nu \beta \beta$ decay.
In this case the mixing elements of $N_1$ in the charged current interaction
are determined by its mass and the Majorana phase of active neutrinos.
We then study the impacts of such a possibility on the direct search for $N_1$. 
In addition, we discuss the consequence of the case
when the $0 \nu \beta \beta$ decay in one specific nucleus is absent.
\end{abstract}
\end{center}
\end{titlepage}

\section{Introduction}
The seesaw mechanism~%
\cite{Minkowski:1977sc,Yanagida:1979as,Yanagida:1980xy,Ramond:1979py,GellMann:1980vs,Glashow:1979nm,Mohapatra:1979ia} 
by right-handed neutrinos with Majorana masses is very attractive because it can provide a
natural explanation of the tiny neutrino masses which have been confirmed
by various oscillation experiments.  
One of the most important consequences is the Majorana nature of 
active neutrinos as well as the heavier states which we call as heavy neutral leptons
(HNLs), and then the lepton number is violated by two units.
In the Standard Model (SM), although the lepton and baryon numbers are accidental symmetries
of the Lagrangian, both are violated by the non-perturbative
quantum effect by the anomaly~\cite{tHooft:1976rip,tHooft:1976snw}.
However, its breaking effect is highly suppressed at zero temperature 
and is essentially negligible in the experimental processes.
Thus, the experimental tests for the lepton number conservation 
are crucially important to verify the seesaw mechanism.

The well-known example of such tests is the search for the neutrinoless 
double beta ($0 \nu \beta \beta$) decay: $(Z,A) \to (Z+2,A)+2 e^{-}$ 
which violates the lepton number by two units 
(see, for example,
reviews~\cite{Doi:1985dx,Pas:2015eia,DellOro:2016tmg,Dolinski:2019nrj}).%
\footnote{
There are other possibilities to test the lepton number violation in
the literature.  They include the inverse $0 \nu \beta \beta$ decay
$e^- e^- \to W^- W^-$~\cite{Rizzo:1982kn,London:1987nz,Dicus:1991fk,Belanger:1995nh,Gluza:1995ky,Gluza:1995ix,Gluza:1995js,Greub:1996ct,Rodejohann:2010jh,Banerjee:2015gca,Asaka:2015oia,Wang:2016eln} and
the rare decays of tau~\cite{Ilakovac:1995km,Ilakovac:1994kj,Ilakovac:1995wc,Gribanov:2001vv,Atre:2005eb} and mesons~\cite{Ng:1978ij,Abad:1984gh,Dib:2000wm,Ali:2001gsa,Atre:2005eb,Asaka:2016rwd}.
}
The current limit of the $0 \nu \beta \beta$ 
decay half-life is $\tau_{1/2} > 1.07 \times 10^{26}$ yr 
by the KamLAND-Zen with $^{136}$Xe~\cite{KamLAND-Zen:2016pfg}.
The decay is mediated by massive active neutrinos if they are Majorana particles,
and their contribution is parameterized by the effective neutrino mass 
$m_{\rm eff}$.  The above half-life limit is translated into the upper bound
on the effective mass as $61-165$~meV~\cite{KamLAND-Zen:2016pfg}. 
Note that the bound on the effective mass recieves
the uncertainty in the nuclear matrix element of
the decay process.


In the seesaw mechanism HNLs 
may participate the $0 \nu \beta \beta$ decay 
which is quantified as an additional part to $m_{\rm eff}$.
When the masses and mixing elements of HNLs are sufficiently light
and large, their contribution to $m_{\rm eff}$ can be comparable to 
that of active neutrinos, and 
the cancellation in $m_{\rm eff}$ between these contributions
happens in some cases.
Notice that the cancellation in $m_{\rm eff}$ 
between three active neutrino is possible 
if the masses of active neutrinos are in the normal hierarchy 
and the lightest active neutrino has a specific mass. 
To achieve this possibility, HNLs must be completely decoupled from the decay. 
In this paper we discuss the seesaw mechanism with two right-handed
neutrinos in which the lightest active neutrino becomes massless.
In this case such a cancellation does not occur and then 
we shall disregard the possibility.

One simple possibility is that all the HNLs participating in the seesaw mechanism
are lighter than about $0.1$~GeV scale (which is the typical momentum scale
in the $0 \nu \beta \beta$ decay).  In this case 
$m_{\rm eff} =0$ is ensured by the intrinsic property in the seesaw mechanism~\cite{Blennow:2010th}.
If this is the case,  the processes of the $0 \nu \beta \beta$
is absent although the lepton number is violated and active neutrinos
induce the sizable contribution to $m_{\rm eff}$.

In a recent article~\cite{Asaka:2020wfo}
we have pointed out another possibility.
It is shown that,
in the minimal choice of the seesaw mechanism with two right-handed neutrinos, 
$m_{\rm eff} = 0$ is realized when 
the heavier HNL decouples from the $0 \nu \beta \beta$ decay but 
the lighter one being lighter than the typical momentum scale of the decay gives a destructive contribution. 
The purpose of this paper is to extend the discussion
to more general cases, especially to the case in which 
the lighter one is heavy as ${\cal O}(1-10)$~GeV scale.
It will be shown that the cancellation in $m_{\rm eff}$ is possible
for such a heavy mass region and the required mixing elements
are relatively large so that such a HNL is a good target for the future search experiments.

The rest of this paper is organized as follow.
In Sec.~\ref{sec:seesaw} we explain the model in the present analysis.
In Sec.~\ref{sec:0nu2beta} we describe the contributions
to the effective neutrino mass in the $0 \nu \beta \beta$ decay from 
active neutrinos as well as HNLs, and then show how the cancellation
in the effective neutrino mass is realized by the lighter HNL. 
In addition, we suggest a possibility that 
even if the $0 \nu \beta \beta$ decay is not observed at an experiment using a specific element, 
other experiments which use different elements can observe the decay 
due to an enhancement originated in the difference of the nuclear matrix elements.
It is then discussed in Sec.~\ref{sec:search} that
the implication of such a cancellation to the direct search of HNLs.
Finally, Sec.~\ref{sec:conclusions} is devoted to discussions and conclusions.
We add Appendices \ref{sec:ap1} and \ref{sec:ap3} to present the physical region of the model parameters 
and the predicted upper and lower bounds of mixing elements 
in each flavor for the HNL $N_1$.

\section{Seesaw model with two right-handed neutrinos}
\label{sec:seesaw}
In this paper we consider the minimal seesaw scenario where 
the SM extended by two right-handed neutrinos $\nu_{R I}$ ($I=1,2$).
The number of right-handed neutrinos must be larger than or equal to
two in order to explain the observed mass squared differences in neutrino oscillations.
The model is described by the Lagrangian 
\begin{align}
  {\cal L}
  =
  {\cal L}_{\rm SM}
  + i \, \overline{\nu_{RI}} \gamma^\mu \partial_\mu \nu_{RI}
  -
  \left(
  F_{\alpha I} \, \overline{\ell_{\alpha}} \, \Phi \, \nu_{RI}
  +
  \frac{M_I}{2} \, \overline{\nu_{RI}^c} \, \nu_{RI}
  +
  h.c.
  \right) \,,\label{Eq:Lag}
\end{align}
Here ${\cal L}_{\rm SM}$ is the SM Lagrangian.  $\Phi$ and
$\ell_\alpha$ ($\alpha = e, \mu,\tau$) are the Higgs and lepton
doublets of the weak SU(2).  Neutrino Yukawa coupling constants and
Majorana masses of right-handed neutrinos are denoted by
$F_{\alpha I}$ and $M_I$, respectively.  Here and hereafter, 
we work in the basis where the mass matrices 
of charged leptons and right-handed neutrinos are diagonal.

When the neutrino masses
of the Dirac type $[M_D]_{\alpha I} = F_{\alpha I} \langle \Phi \rangle$
is much smaller than the Majorana masses $M_I$, 
{\it i.e.}, $|[M_D]_{\alpha I}|
\ll M_I$, the seesaw mechanism is realized, 
and the mass matrix of active neutrinos $\nu_i$ ($i=1,2,3$)
is given by 
\begin{align}
  [M_\nu]_{\alpha \beta} = - \frac{[M_D]_{\alpha I} [M_D^T]_{I \beta}}{M_I} \,, 
  \label{Eq:M_nu}
\end{align}
which is diagonalized by the neutrino mixing matrix $U$ 
called as the PMNS matrix~\cite{Pontecorvo:1957qd,Maki:1962mu}. 
The remaining mass eigenstates are HNLs denoted by 
$N_I$ which almost correspond to right-handed neutrino states.
The seesaw mechanism tells that the mass of $N_I$ is given by $M_I$%
\footnote{Throughout this paper, we neglect higher order corrections 
in the expansion by $[M_D]_{\alpha I}/M_I$ to both active neutrino and HNL masses.}. 
These neutrino mass eigenstates take part in the weak gauge interactions through the mixing effect as 
\begin{align}
  \nu_{L \alpha} = U_{\alpha i} \, \nu_i + \Theta_{\alpha I} N_I^c \,,
\end{align}
where the mixing elements of $N_I$ is given by
$\Theta_{\alpha I} = [M_D]_{\alpha I} \, M_I^{-1}$.

We apply the parametrization of the couplings
by Casas and Ibarra~\cite{Casas:2001sr,Abada:2006ea} as
\begin{eqnarray}
  \label{eq:F}
    F = \frac{i}{\vev{\Phi}} \,
    U \, D_\nu^{1/2} \, \Omega \, D_N^{1/2} \,,
\end{eqnarray}
where $D_\nu = \mbox{diag}(m_1, m_2, m_3)$.
In the considering case with two right-handed neutrinos
the lightest active neutrino becomes massless and then
the possible mass orderings are
$m_3 > m_2 > m_1 = 0$ for the normal hierarchy (NH) case 
and
$m_2 > m_1 > m_3 = 0$ for the inverted hierarchy (IH) case,
respectively.
The mixing matrix of active neutrinos is expressed as
\begin{eqnarray}
  U = 
  \left( 
    \begin{array}{c c c}
      c_{12} c_{13} &
      s_{12} c_{13} &
      s_{13} e^{- i \delta} 
      \\
      - c_{23} s_{12} - s_{23} c_{12} s_{13} e^{i \delta} &
      c_{23} c_{12} - s_{23} s_{12} s_{13} e^{i \delta} &
      s_{23} c_{13} 
      \\
      s_{23} s_{12} - c_{23} c_{12} s_{13} e^{i \delta} &
      - s_{23} c_{12} - c_{23} s_{12} s_{13} e^{i \delta} &
      c_{23} c_{13}
    \end{array}
  \right)  
  \times
  \mbox{diag} 
  ( 1 \,,~ e^{i \eta} \,,~ 1) \,,
\end{eqnarray}
with $s_{ij} = \sin \theta_{ij}$ and $c_{ij} = \cos \theta_{ij}$.  
$\delta$ and $\eta$ are the Dirac and Majorana phases, respectively.
$D_N = \mbox{diag}(M_1,M_2)$ is the mass matrix of HNLs.
The $3 \times 2$ matrix $\Omega$ can be taken in the form
\begin{eqnarray}
  \Omega =
  \left(
    \begin{array}{c c}
      0 & 0 \\
      \cos \omega & - \sin \omega \\
      \xi \sin \omega & \xi \cos \omega
    \end{array}
  \right) ~~~\mbox{for the NH case} \,,~~~~
  \left(
    \begin{array}{c c}
      \cos \omega & - \sin \omega \\
      \xi \sin \omega & \xi \cos \omega \\
      0 & 0 
    \end{array}
  \right) ~~~\mbox{for the IH case} \,,
\end{eqnarray} 
where $\xi = \pm 1$ is  sign parameter 
and $\omega$ is a complex parameter
\begin{align}
	\omega = \omega_r + i \omega_i \,,
\end{align}
with $\omega_r$ ($\omega_i$) as the real (imaginary) part of $\omega$.
In Eq.~(\ref{eq:F}) 
the masses and mixing angles of active neutrinos, which are relevant 
for the neutrino oscillations, are
automatically satisfied reproducing the experimental results 
being independent of the choice of $D_N$ and $\Omega$.  
Further, it is convenient to introduce
\begin{align}
    X_\omega = \exp ( \omega_i ) \,, 
\end{align}
because it represents the overall strength of the Yukawa coupling matrix 
(see, for example, the discussion in Ref.~\cite{Asaka:2011pb}). 
In practice the Yukawa coupling constants can be enhanced by taking
large values of $|\omega_i|$, as $F \propto X_\omega$
or $X_\omega^{-1}$ for $X_\omega \gg 1$ or $\ll 1$. 
As a result, the enhancement of the Yukawa coupling constants is reflected to 
the enhancement of the mixing elements $\Theta_{\alpha I}$. 

In the present analysis we adopt the convention
where the ranges of the parameters 
are $\omega_r \in [ - \frac{\pi}{2}, \frac{\pi}{2}]$ 
and $X_\omega \in [0, + \infty]$%
\footnote{Technically speaking, $X_\omega$ (or $\omega_i$) can take the value 
in the whole range $[0,+\infty]$, 
however, the region is restricted not to exceed 
the limits of the mixing elements from the direct search experiments and 
the perturbative limit 
of the neutrino Yukawa coupling constants in practice.} 
with fixing $\xi = + 1$ 
(see the discussion in Appendix~\ref{sec:ap1}).  
Note that the Majorana phase takes a value in $\eta \in [0, \pi]$. 

\section{Neutrinoless double beta decay}
\label{sec:0nu2beta}
Now let us discuss the $0 \nu \beta \beta$ decay
in the model under consideration.
The half-life of the decay is parameterized as
\begin{align}
	\tau_{1/2}^{-1} = G  \, |{\cal M} m_{\rm eff} |^2 \,,
\end{align}
where $G$ is the phase-space factor, ${\cal M}$ is the nuclear matrix element of active neutrinos 
and $m_{\rm eff}$ is the effective neutrino mass in the $0 \nu \beta \beta$ decay. 
Since all Majorana neutrinos, {\it i.e.},
not only active neutrinos but also HNLs,
contribute to the decay, 
the effective mass is given by 
\begin{align}
	m_{\rm eff} =
	m_{\rm eff}^\nu + m_{\rm eff}^{N_1} + m_{\rm eff}^{N_2} \,.\label{Eq:general-meff}
\end{align}
Note that we consider $m_{\rm eff}$ as a complex number 
while the observed value has to be its absolute value.
The contribution from active neutrinos is
\begin{align}
	m_{\rm eff}^\nu = \sum_i \, U_{ei}^2 \, m_i \,.
\end{align}
By taking the central values 
of the mass squared differences, the mixing angles and the Dirac phase in the PMNS matrix given in Ref.~\cite{nufit} 
and by varying the Majorana phase $\eta$
we find that 
$|m_{\rm eff}^\nu| = 1.45$--$3.68$~meV
and 18.6--48.4~meV for the NH and IH cases, respectively.
On the other hand, the effective mass from each HNL is
\begin{align}
	m_{\rm eff}^{N_I} &= \Theta_{e I}^2 \, M_I \, f_\beta (M_I) \,,
\end{align}
which can be written explicitly as 
\begin{align}
	m_{\rm eff}^{N_1} &= \Theta_{e 1}^2 \, M_1 \, f_\beta (M_1) 
	= 
	\left\{
		\begin{array}{ll}
			- \left( U_{e2} m_2^{1/2} \cos \omega + U_{e3} m_3^{1/2} \sin \omega \right)^2
			f_\beta (M_1)
			&
			\mbox{for the NH case}
			\\[1ex]
			- \left( U_{e1} m_1^{1/2} \cos \omega + U_{e2} m_2^{1/2} \sin \omega \right)^2
			f_\beta (M_1)
			&
			\mbox{for the IH case}
		\end{array}
	\right.\,,\\[2ex]
	m_{\rm eff}^{N_2} &= \Theta_{e 2}^2 \, M_2 \, f_\beta (M_2) 
	= 
	\left\{
		\begin{array}{ll}
			- \left( U_{e2} m_2^{1/2} \sin \omega - U_{e3} m_3^{1/2} \cos \omega \right)^2
			f_\beta (M_2)
			&
			\mbox{for the NH case}
			\\[1ex]
			- \left( U_{e1} m_1^{1/2} \sin \omega - U_{e2} m_2^{1/2} \cos \omega \right)^2
			f_\beta (M_2)
			&
			\mbox{for the IH case}
		\end{array}
	\right.
	\,,
\end{align}
where $f_\beta (M_I)$ represents the suppression 
in the nuclear matrix element by the propagator of $N_I$.
Here we follow the results in Refs.~\cite{Faessler:2014kka,Barea:2015zfa}
and take the form 
\begin{align}
	f_\beta (M_I) = \frac{\Lambda_\beta^2}{\Lambda_\beta^2 + M_I^2} \,,\label{Eq:fbeta}
\end{align}
where $\Lambda_\beta$ is the typical scale of the Fermi momentum
and we take $\Lambda_\beta = 200$~MeV as a reference value throughout this analysis. 
We will discuss the impact of the change of the Fermi momentum later.

It is known that $m_{\rm eff}=0$ is possible even if $m_{\rm eff}^\nu \neq 0$ 
due to the presence of HNLs participating the seesaw mechanism. 
Namely, when both $N_1$ and $N_2$ are sufficiently lighter than $\Lambda_\beta$, 
$f_\beta (M_{1,2}) = 1$ and then the seesaw mechanism 
guarantees the following equality~\cite{Blennow:2010th} 
\begin{align}
	m_{\rm eff} = \sum_i U_{ei}^2 \, m_i + \sum_I \Theta_{eI}^2 \, M_I = 0 \,.
\end{align}
Thus, even though the lepton number is violated by the Majorana masses
of right-handed neutrinos, the $0 \nu \beta \beta$ decay processes are absent 
being independent of how large $\Lambda_\beta$ is.

We have pointed out in Ref.~\cite{Asaka:2020wfo} another possibility in the case 
when the masses of $N_1$ and $N_2$ are hierarchical.  
It is assumed that the mass and mixing elements of $N_2$ are sufficiently heavy and small 
so that $m_{\rm eff}^{N_2}$ can be neglected. 
In that analysis we have also assumed that $M_1 \ll \Lambda_\beta$
which gives $f_\beta(M_1) = 1$ approximately. 
The effective mass in this case is abridged as
\begin{align}
	m_{\rm eff}
	=
	\left\{
		\begin{array}{l l}
			\left( U_{e2} m_2^{1/2} \sin \omega - U_{e3} m_3^{1/2} \cos \omega
			\right)^2 &~~~\mbox{for the NH case}
			\\[2ex]
			\left( U_{e1} m_1^{1/2} \sin \omega - U_{e2} m_2^{1/2} \cos \omega
			\right)^2 &~~~\mbox{for the IH case}			
		\end{array}
	\right. \,.
\end{align}
It is found that $m_{\rm eff}$ vanishes if the complex parameter $\omega$ satisfies
\begin{align}
	\label{eq:A}
	\tan \omega 
	=
	A
	=
	\left\{
		\begin{array}{l l}
			\displaystyle
			\frac{U_{e3} m_3^{1/2}}{U_{e2} m_2^{1/2} }
			&
			~~~\mbox{for the NH case}
			\\[3ex]
			\displaystyle
			\frac{U_{e2} m_2^{1/2}}{U_{e1} m_1^{1/2} }
			&
			~~~\mbox{for the IH case} 
		\end{array}
	\right. \,.
\end{align}
As described in Ref.~\cite{Asaka:2020wfo},
the mixing elements of $N_1$, $\Theta_{\alpha 1}$, 
are determined from the mass $M_1$ 
and the Majorana phase $\eta$ of active neutrinos in this situation 
and the flavor structure among them is different in each mass hierarchy of active neutrinos.  
Then, the relative sizes of $\Theta_{\alpha 1}$, 
if they will be measured in future experiments, 
give the important information of $\eta$ 
which is not basically determined by neutrino oscillation experiments 
as well as the mass hierarchy of active neutrinos.

In this paper we would like to extend the above discussions to more general cases.
We assume again $f_\beta (M_2) =0$~%
\footnote{
Our results do not change much as long as the contribution from $N_2$ 
is sufficiently suppressed.
}
, but consider the case $f_\beta (M_1) \neq 1$ which is valid for 
heavier $N_1$ with $M_1 \gtrsim \Lambda_\beta$.  
Hereafter we present the analytic results in the NH case of active neutrino masses 
and the extension to the IH case is straightforward.
In this case, the effective mass can be written as
\begin{align}
	m_{\rm eff}
	= \left( U_{e2} m_2^{1/2} \sin \omega - U_{e3} m_3^{1/2} \cos \omega \right)^2
	+ \left( U_{e2} m_2^{1/2} \cos \omega + U_{e3} m_3^{1/2} \sin \omega \right)^2
	\times \delta_f^2 \,,	
\end{align}
where we have introduced $\delta_f$ ($1 > \delta_f>0$) as
\begin{align}
	f_\beta (M_1) = 1 - \delta_f^2 \,.
\end{align}
It is then found that the effective mass vanishes if
\begin{align}
	\label{eq:TanOM}
	\tan \omega = \frac{ A \pm i \delta_f}{1 \mp i \delta_f A} 
	\equiv \tan \omega_\pm\,.
\end{align}
Note that this condition gives Eq.~(\ref{eq:A}) 
for $\delta_f \to 0$, as it should be.
This result itself can be applied to both mass hierarchies 
by taking proper $A$ defined in Eq.~(\ref{eq:A}).

\begin{figure}[t]
  \centerline{
  \includegraphics[width=7cm]{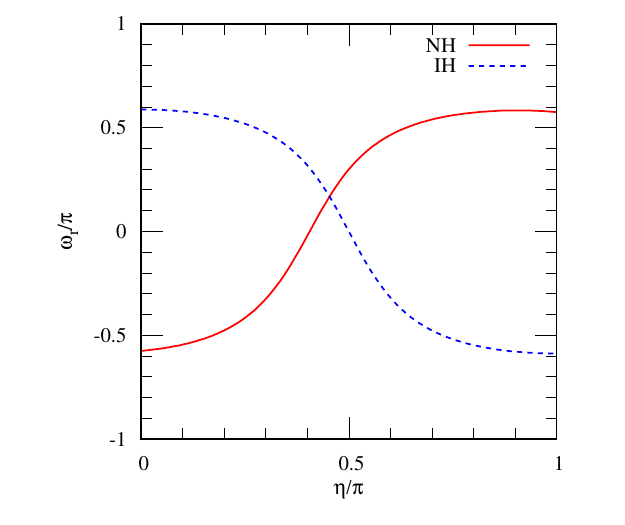}%
  }%
  \vspace{-2ex}
  \caption{
  	The required value of $\omega_r$ 
	  for the vanishing effective neutrino mass
		in the NH (red-solid line) or IH (blue-dashed line) case.
  }
  \label{FIG_omr}
\end{figure}
Interestingly, we find that the required value for the real part $\omega_r$ 
is independent of $\delta_f$ (\ie, the mass of $N_1$): 
\begin{align}
	\tan 2 \omega_r
	=
	\frac{2 \mbox{Re}A}{1 - |A|^2} \,.\label{Eq:wrconst}
\end{align}
This relation holds for both signs in Eq.~(\ref{eq:TanOM}).
Since it can be applied to the case 
$\delta_f = 0$ (\ie, $f_\beta (M_1) = 1$),
the result concerning on $\omega_r$ in Ref.~\cite{Asaka:2020wfo} 
is substantiated even in this case.  
The required value of $\omega_r$ is determined by the Majorana phase, 
which is shown in Fig.~\ref{FIG_omr}. 
On the other hand, the required value of $\omega_i$ is
represented by using $X_\omega$ as
\begin{align}
	\label{eq:Xom}
	X_\omega^2
	=
	\frac{1 \pm \delta_f}{1 \mp \delta_f}
	\sqrt{ \frac{ 1 + |A|^2 + 2 \mbox{Im}A }{ 1 + |A|^2 - 2 \mbox{Im}A } } \,,
\end{align}
where the upper/lower sign corresponds to that in Eq.~(\ref{eq:TanOM}).

\begin{figure}[t]
  \centerline{
  \includegraphics[width=7cm]{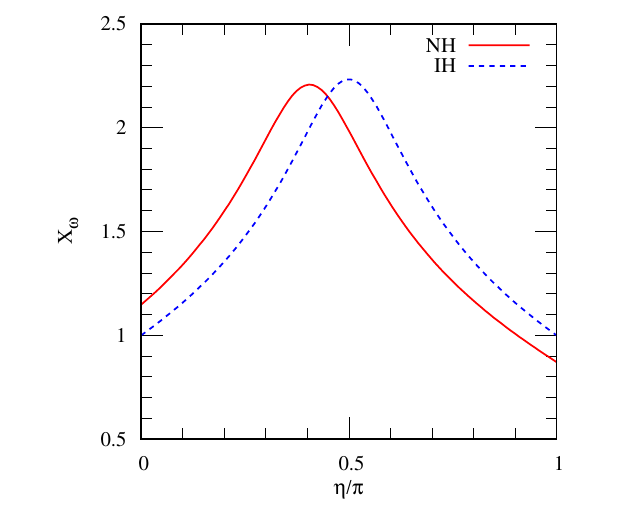}%
  }%
  \vspace{-2ex}
  \caption{
  	The required value of $X_\omega$ 
	  for the vanishing effective neutrino mass
		in the NH (red-solid line) or IH (blue-dashed line) case.
		Here we take $\delta_f =0$.
  }
  \label{FIG_Xom_f=1}
\end{figure}

When $\delta_f \to 0$ (\ie\,, $f_\beta (M_1) \to 1$),
it becomes independent of $M_1$ and is determined by the Majorana phase 
and turns out to be the results in Ref.~\cite{Asaka:2020wfo}.
This point is shown in Fig.~\ref{FIG_Xom_f=1}.
It is seen that the required value $X_\omega$ is of the order of the unity
and it takes the maximal value at $\eta = 0.41 \pi$ or 
$\pi/2$ in the NH or IH case, respectively. 
This can simply be read from the dependence on the CP phases 
$\delta$ and $\eta$ in each mass hierarchy. 
We find a simple relationship between $\omega$ and the CP phases 
in the matrix $U$ under the cancellation condition 
\begin{align}
\frac{\sinh 2 \omega_i}{\sin 2 \omega_r} 
=
\frac{1}{2} \left( X_\omega^2 - X_\omega^{-2} \right) \frac{1}{\sin 2 \omega_r}
= 
\frac{{\rm Im} A}{{\rm Re} A}
	=
	\left\{
		\begin{array}{l l}
			\displaystyle
			- \tan (\delta + \eta)
			&
			~~~\mbox{for the NH case}
			\\[3ex]
			\displaystyle
			\tan \eta
			&
			~~~\mbox{for the IH case} 
		\end{array}
	\right.\,.\label{Eq:PH-L}
\end{align}

\begin{figure}[t]
  \centerline{
  \includegraphics[width=7cm]{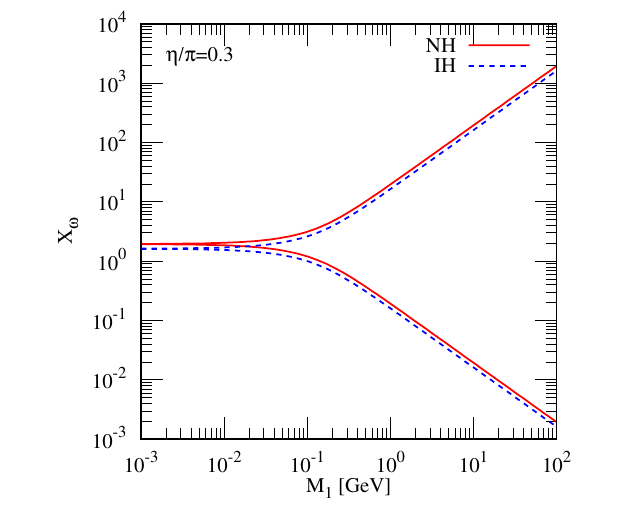}%
  }%
  \vspace{-2ex}
  \caption{
  	The required value of $X_\omega$ 
	  for the vanishing effective neutrino mass
		in the NH (red-solid line) or IH (blue-dashed line) case.
		Here we take the Majorana phase $\eta = 0.3 \pi$.
  }
  \label{FIG_Xom_E03}
\end{figure}

When $\delta_f \neq 0$, there are two possibilities
for $X_\omega$ depending on the sign in Eq.~(\ref{eq:Xom}).
The result with $\eta = 0.3 \pi$ is shown in Fig.~\ref{FIG_Xom_E03}.
It is seen that, when $M_1$ becomes smaller than $\Lambda_\beta$,
$X_\omega$ approaches to the value shown in Fig.~\ref{FIG_Xom_f=1}.
On the other hand, as $M_1 \gg \Lambda_\beta$,
the required value of $X_\omega$ is proportional to 
$M_1$ or $M_1^{-1}$ for the upper or lower sign in Eq.~(\ref{eq:Xom}). 
This behavior can be understood from the fact that 
$\delta_f$ approaches to zero as $M_1$ goes smaller enough than $\Lambda_\beta$ 
and monotonically depends on $M_1$ as it gets larger. 
This is one of the important results in the present analysis, especially
when we discuss the impacts on the direct search of $N_1$ 
in the next section.  
This is because 
$X_\omega$ (or $X_\omega^{-1}$) sets the overall scale of the mixing
elements of HNLs.  See, for example, the discussion in Ref.~\cite{Asaka:2011pb}.

As similar equation to Eq.~(\ref{Eq:PH-L}), 
we can also get the simple relationship even when $\delta_f \neq 0$ as 
\begin{align}
    \frac{\sinh 2 \omega_i}{\sin 2 \omega_r} 
    =
    \frac{1}{2} \left( X_\omega^2 - X_\omega^{-2} \right) \frac{1}{\sin 2 \omega_r}
    =
    \frac{(1+\delta_f^2) {\rm Im} A \pm \delta_f (1+|A|^2)}{(1-\delta_f^2) {\rm Re} A
    }\,,\label{Eq:PH-H}
\end{align}
which can be applied to both NH and IH cases depending on the choice of $A$. 
By using this expression, we can get 
\begin{align}
    X_\omega^2 
    = 
    \zeta 
    +
    \left( 
    \zeta^2 
    + 1
    \right)^{1/2}\,,\label{Eq:Xome-zeta}
\end{align}
where $\zeta$ is defined as 
\begin{align}
    \zeta 
    = 
    2 \frac{(1+\delta_f^2) {\rm Im} A \pm \delta_f (1+|A|^2)}{(1-\delta_f^2)\sqrt{(1-|A|^2)^2 + 4 {\rm Re} A^2}}\,,
\end{align}
which is consistent with Eq.~(\ref{eq:Xom}). 

\begin{figure}[t]
  \centerline{
  \includegraphics[width=7cm]{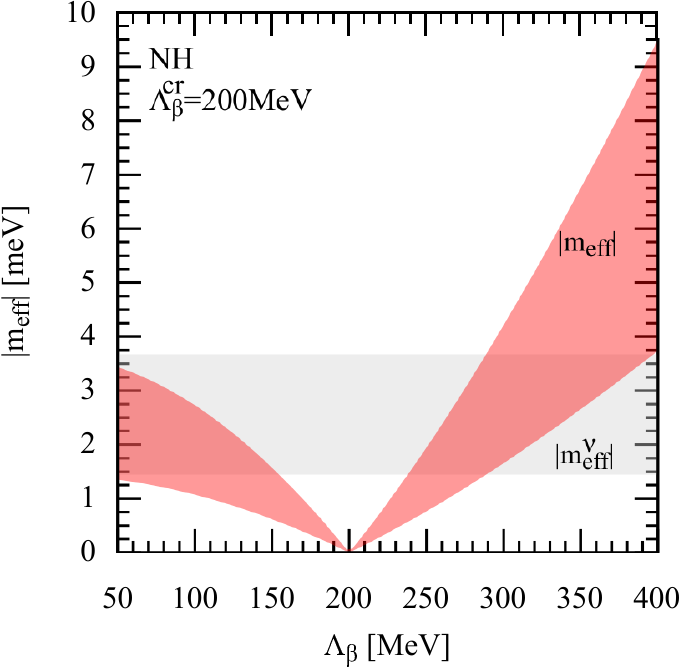}%
  \includegraphics[width=7.2cm]{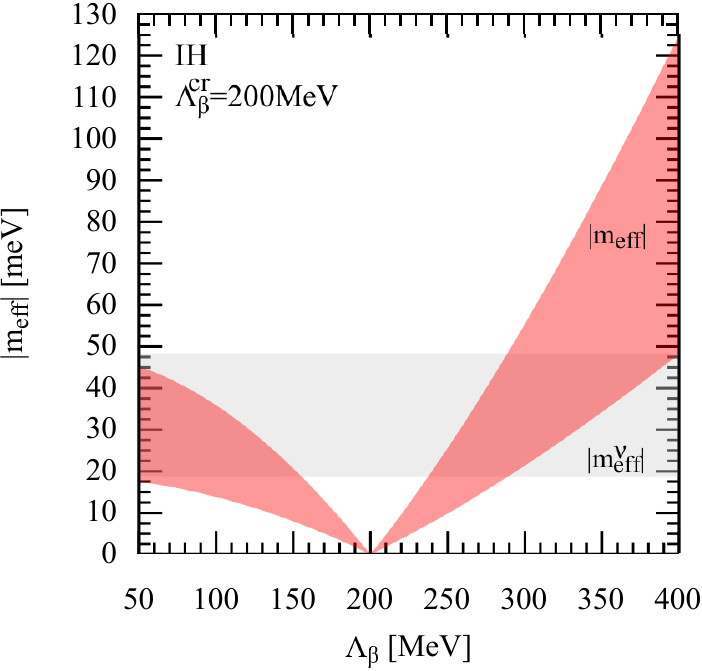}%
  }%
  \vspace{-2ex}
  \caption{
  The range of the effective mass $|m_{\rm eff}|$ (red-region)
  when the $0 \nu \beta \beta$ decay of a nucleus 
  with $\Lambda_\beta = \Lambda_\beta^{\rm cr} =200$~MeV
  is absent.  The range of the effective mass from active neutrinos
  $|m_{\rm eff}^\nu|$ (gray-region) is also shown.
  Here we take $M_1=1$~GeV and $M_2=200$~GeV.
  }
  \label{FIG_Meff_Ldep}
\end{figure}
Before closing this section, we should mention the form 
of the function $f_\beta$, the choice of the scale $\Lambda_\beta$ 
and its impacts on the observations. 
These quantities must be evaluated by the detailed calculation
of the nuclear matrix element for a given $0 \nu \beta \beta$
decay nucleus.  Especially, the value of $\Lambda_\beta$
varies depending on the nucleus of each experiments 
and receives an uncertainty of the nuclear physics~
(see, for example, Refs.~\cite{Faessler:2014kka,Barea:2015zfa}).

As for the case when $M_1 \ll \Lambda_\beta$, 
the cancellation (or a partial cancellation) occurs
universally since $f_\beta (M_1) \simeq 1$
for all the $0 \nu \beta \beta$ nuclei.
In this case, thus, if the decay is absent for a specific nucleus,
the same thing happens for other nuclei.
This is the main observation in Ref.~\cite{Asaka:2020wfo}.

On the other hand, when $M_1 \gtrsim \Lambda_\beta$, 
the situation is more involved. 
Let consider the case when the cancellation in $m_{\rm eff}$ occurs
as described above for one nucleus with 
$\Lambda_\beta = \Lambda_\beta^{\rm cr}$ 
which we have take to be $200~{\rm MeV}$ as a reference value 
in our main discussions.
Then, the rates of the $0 \nu \beta \beta$ decays
for other nuclei can be suppressed or even enhanced 
depending on the value of $\Lambda_\beta$ for the nuclei~%
\footnote{This point has been discussed in earlier papers~\cite{Halprin:1983ez,Leung:1984vy} 
in the context of one light and one heavy Majorana neutrinos. 
Here we consider the case which is consistent with the oscillation experiments 
in the minimal seesaw mechanism.}. 
To see this, by denoting 
$f_\beta^{\rm cr} (M_1) = f_\beta(M_1)|_{\Lambda_\beta = \Lambda_\beta^{\rm cr}}$, 
the cancellation condition gives us that 
\begin{align}
    M_1 \Theta_{e1}^2 = - \frac{m_{\rm eff}^\nu}{f_\beta^{\rm cr}(M_1)}\,.
\end{align}
By inserting this expression into Eq.(\ref{Eq:general-meff}) 
with dropping $N_2$ contribution, 
we get
\begin{align}
    m_{\rm eff} = m_{\rm eff}^\nu \left( 1 - \frac{f_\beta (M_1)}{f_\beta^{\rm cr}(M_1)} \right)\,,
\end{align}
where $f_\beta (M_1)$ can be different from the critical value depending on the elements. 
Namely, if $f_\beta (M_1)$ becomes bigger than $f_\beta^{\rm cr}$, 
which corresponds to $\Lambda_\beta \gtrsim \sqrt{2} \Lambda_\beta^{\rm cr}$, 
the predicted effective mass can be greater than $|m_{\rm eff}^\nu|$ at another experiment. 
As shown in Fig.~\ref{FIG_Meff_Ldep}, 
the predicted effective mass can overcome $|m_{\rm eff}^\nu|$ 
when $\Lambda_\beta \gtrsim 290~{\rm MeV}$ in the both mass hierarchy.

\section{Impacts on search for heavy neutrino}
\label{sec:search}
Next, we turn to discuss the consequences of the no $0 \nu \beta \beta$ decay
due to the destructive contribution of $N_1$.
Since the complex parameter $\omega$ is fixed as shown in Eq.~(\ref{eq:TanOM}),
the mixing elements of $N_1$, $\Theta_{\alpha 1}$, are predicted by its mass and the Majorana phase
when we use the central values of mixing angles, mass squared differences
and the Dirac phase from the oscillation experiments~\cite{nufit}.

\begin{figure}[t]
  \centerline{
  \includegraphics[width=7cm]{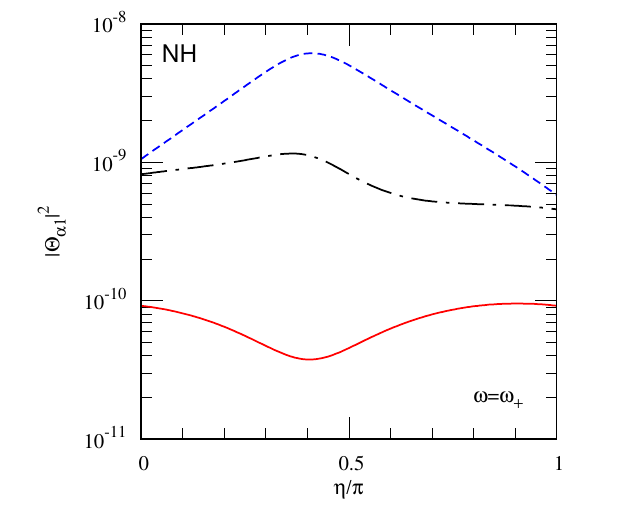}
  \includegraphics[width=7cm]{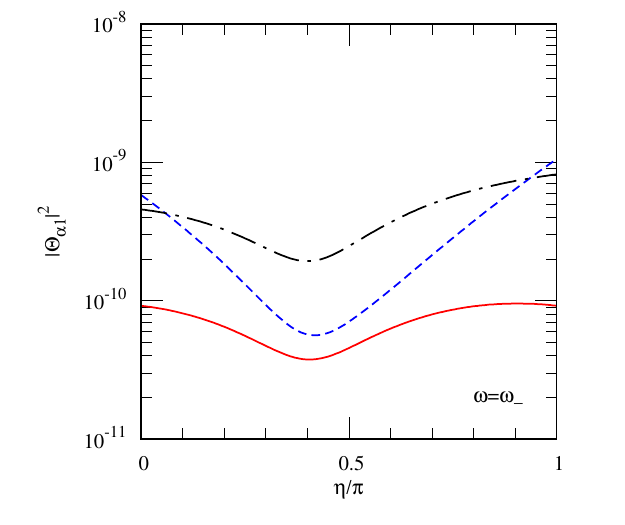}
  }
  \centerline{
  \includegraphics[width=7cm]{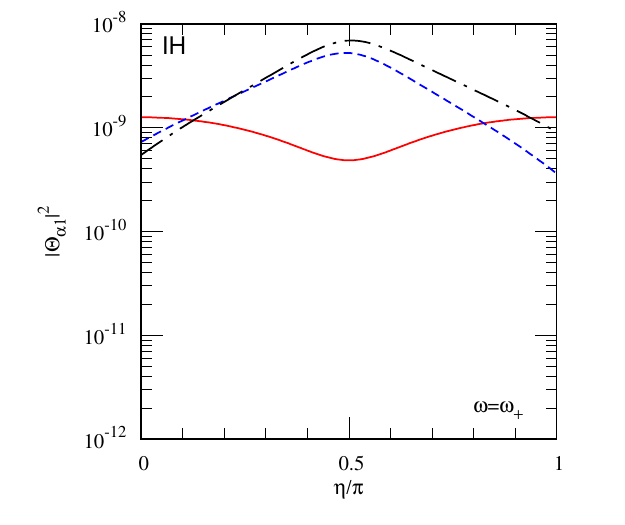}
  \includegraphics[width=7cm]{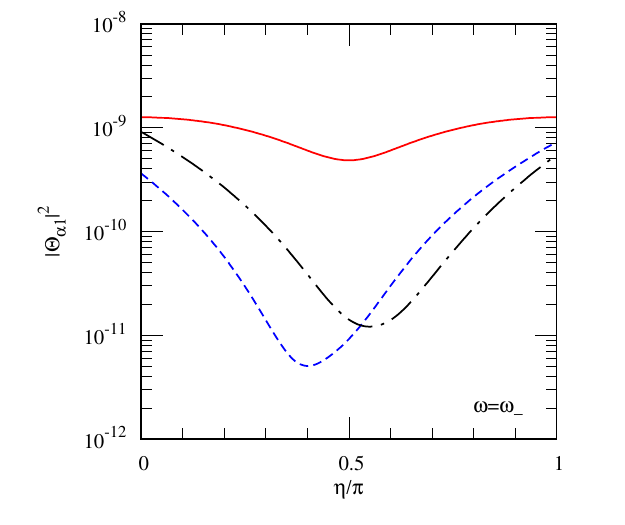}
  }
  \vspace{-2ex}
  \caption{
  	The mixing elements $|\Theta_{\alpha 1}|^2$  
	  for the vanishing effective neutrino mass
		in the NH (upper panel) or IH (lower panel).
		The left or right panel is for 
		the case with $\omega = \omega_+$ or $\omega = \omega_-$ 		
		in Eq.~(\ref{eq:TanOM}), respectively.
		Here we take the $N_1$ mass $M_1 =1$~GeV.
		}
  \label{FIG_THsq_M1}
\end{figure}

Especially, among all flavor mixing elements, the electron-type mixing element is simply given by
\begin{align}
	\label{eq:THsqe1}
	| \Theta_{e1}|^2 = \frac{| m_{\rm eff}^\nu |}{ M_1 f_\beta (M_1) } \,,
\end{align}
which is a direct consequence of $m_{\rm eff} =0$.
Note that $|\Theta_{e1}|^2$ is independent of 
the choice of $\omega = \omega_+$ or $\omega_-$
and is uniquely determined solely by the mass $M_1$ 
as well as active neutrino parameters in $m_{\rm eff}^\nu$.
On the other hand, the elements with $\omega = \omega_+$ are 
larger than those with $\omega = \omega_-$ for $|\Theta_{\mu 1}|^2$
and $|\Theta_{\tau 1}|^2$ in the wide region of $\eta$ as shown in Fig.~\ref{FIG_THsq_M1} 
where we fix $M_1 = 1~{\rm GeV}$. 
This feature is almost independent of the choice of $M_1$. 
Depending on the choice of $\omega$, 
the muon- and tau-type elements are very different, 
namely whether making a peak or a bump, 
which is helpful to identify the value of $X_\omega$ 
(\ie\,, the imaginary part of $\omega$).

Further, the relative sizes of the mixing elements are very sensitive to 
the Majorana phase $\eta$.
In Fig.~\ref{FIG_THsq_M1} we show the relative size of each mixing elements 
in terms of the Majorana phase. 
In the NH case, the electron-type mixing element is always smallest among all flavors, 
in the IH case, on the other hand, there is some possibilities 
where the electron-type element can be the largest when $\omega=\omega_+$ 
and further it is always largest when $\omega=\omega_-$. 
In addition to that, the order between muon- and tau-type elements depends on $\eta$. 
When $\omega=\omega_+$ in the NH case, 
the electron element becomes far below than others. 
When $\omega=\omega_-$ in the NH case, 
the magnitude of all flavor mixing elements gets closer 
but the flavor structure is different from those of $\omega_+$ in wide region of $\eta$. 
On the other hand, in the IH case, 
$\omega=\omega_+$ gives degenerated magnitude for all flavor mixing elements 
and relatively stronger than any other cases. 
When $\omega=\omega_-$ in the IH case, 
other than the electron mixing element become 
far below than the electron mixing elements by taking specific $\eta$. 
Therefore, muon- and tau-type mixing elements have feebly chance to be detected.

\begin{figure}[t]
  \centerline{
  \includegraphics[width=7cm]{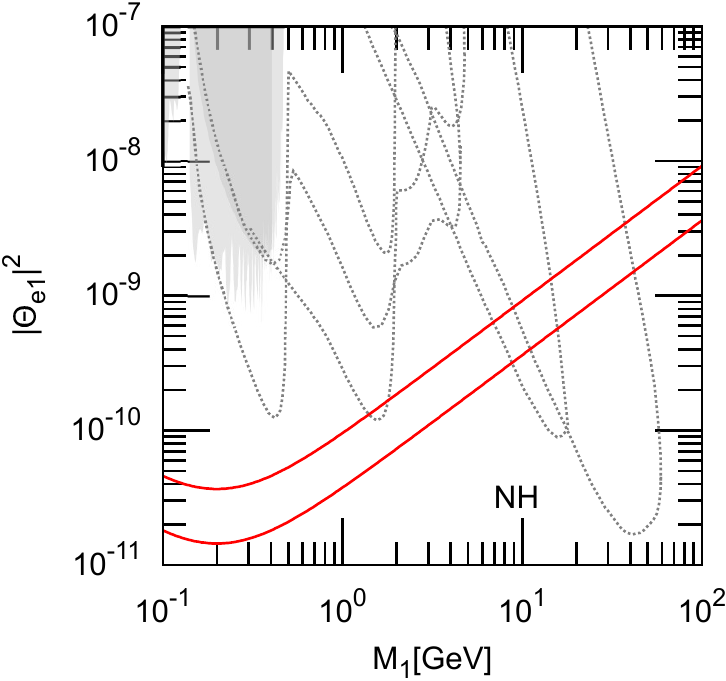}
  \includegraphics[width=7cm]{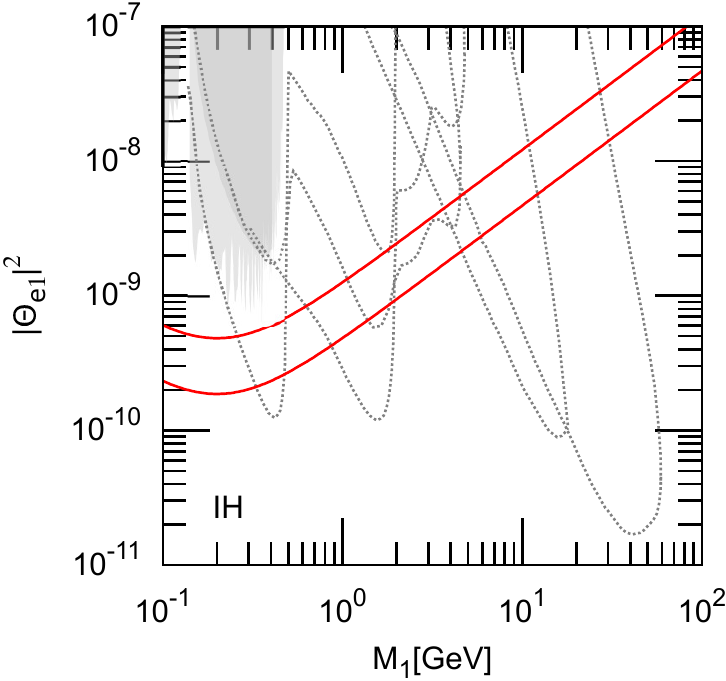}%
  }%
  \vspace{-2ex}
  \caption{
  {
    The region of the mixing element $|\Theta_{e 1}|^2$  
	  for the vanishing effective neutrino mass
	  (between two red lines)
		in the NH (left panel) or IH (right panel) case.
		Here we vary the Majorana phase $\eta = 0$ to 
		$\pi$.
		The shaded regions are excluded by the direct searches for HNL.
		The dotted lines shows the sensitivities on $|\Theta_{e 1}|^2$ 
		by future search experiments.
		}
  }
  \label{FIG_THsq_E03}
\end{figure}

In Fig.~\ref{FIG_THsq_E03} we show the range of the mixing element 
 $|\Theta_{e1}|^2$ by varying the Majorana phase from $\eta = 0$ to $\pi$ in terms of
$M_1$.
The dependence on $M_1$ drastically changes at around $M_1 = \Lambda_\beta$ 
correlating with Eq.~(\ref{Eq:fbeta}). 
Namely, since $M_1$ gets exceed $\Lambda_\beta$, 
$f_\beta$ works as a suppression factor, 
the mixing element $|\Theta_{e 1}|^2$ has to become larger (by enlarging $X_\omega$ or $X_\omega^{-1}$) 
to realize the cancellation of the effective mass. 
This feature is advantageous for the direct search experiments. 
In Fig.~\ref{FIG_THsq_E03} we also show the current bounds 
from various searches~\cite{PIENU:2011aa,Aguilar-Arevalo:2017vlf,NA62:2020mcv} 
and also the sensitivities by the future experiments~%
\cite{Blondel:2014bra,SHiP:2018xqw,Krasnov:2019kdc,Alpigiani:2020tva}. 
It is seen that a wide range of $|\Theta_{e1}|^2$ can be probed by the future experiments, especially for the IH case.
On the other hand, the results of other elements,
$|\Theta_{\mu 1}|^2$ and $|\Theta_{\tau 1}|^2$,
are shown in Appendix~\ref{sec:ap3}.


\begin{figure}[ht]
  \centerline{
  \includegraphics[width=7cm]{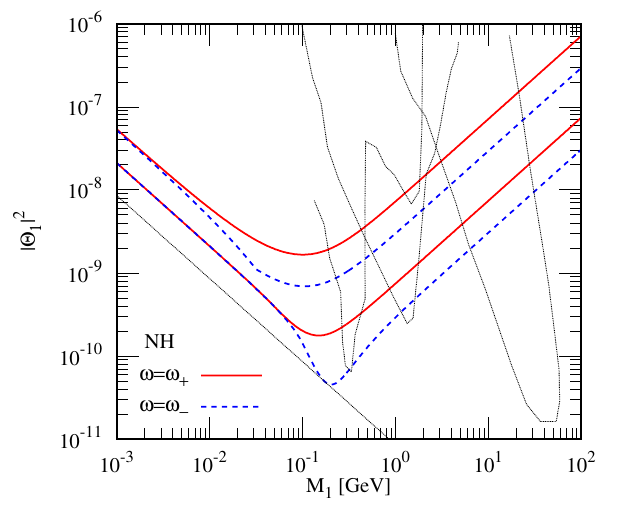}
  \includegraphics[width=7cm]{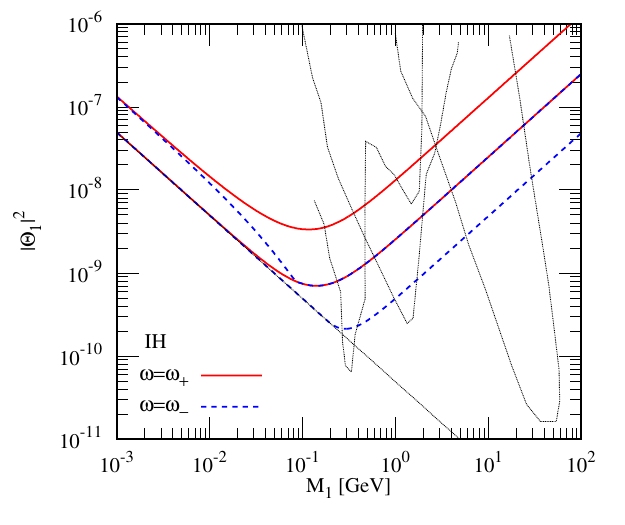}
  }

  \vspace{-2ex}
  \caption{
  	The maximal and minimal values of the mixing elements $|\Theta_{1}|^2$  
	  for the vanishing effective neutrino mass
	  by varying the Majorana phase.
	  The left or right panel is for the NH or IH case.
	  The red-sold or blue-dashed lines are for the case 
	  with $\omega = \omega_+$ or $\omega_-$
      in Eq.~(\ref{eq:TanOM}).
  }
  \label{FIG_THsq_OMpm}
\end{figure}


Furthermore, the sum of the $N_1$ mixing elements is given by
\begin{align}
	| \Theta_{1} |^2
	= \sum_\alpha |\Theta_{\alpha 1}|^2
	= 
	\left\{
	\begin{array}{ll}
		\displaystyle
		\frac{1}{M_1} 
		\left[ 
		\frac{m_3 + m_2}{4} \left( X_\omega^2 + X_\omega^{-2} \right)
   	- \frac{m_3 - m_2}{2} \cos ( 2 \omega_r) 
   	\right]
	  &
	  \mbox{for the NH case} 
	  \\[2ex]
	  \displaystyle
		\frac{1}{M_1} 
		\left[ 
		\frac{m_2 + m_1}{4} \left( X_\omega^2 + X_\omega^{-2} \right)
   	- \frac{m_2 - m_1}{2} \cos ( 2 \omega_r) 
   	\right]
	  &
	  \mbox{for the IH case} 
	\end{array}
	\right.
	\,.\label{Eq:SUM-SQ}
\end{align}
Since the cancellation in $m_{\rm eff}$
requires the specific values of $\omega_r$ and $X_\omega$ as
shown in Eqs.~(\ref{Eq:wrconst}) and (\ref{Eq:Xome-zeta}),
we obtain 
\begin{align}
    | \Theta_{1} |^2
	=
	\left\{
	\begin{array}{ll}
    \displaystyle
		\frac{1}{M_1} 
		\left[ 
		\frac{m_3 + m_2}{2} 
		\left( 
    \zeta^2 
    + 1
    \right)^{1/2}
   	\pm \frac{m_3 - m_2}{2} 
   	    \frac{1-|A|^2}{\sqrt{(1-|A|^2)^2 + 4 {\rm Re} A^2}}
   	\right]
   	& \mbox{for the NH case} 
   	\\[4ex]
   	\displaystyle
   	\frac{1}{M_1} 
		\left[ 
		\frac{m_2 + m_1}{2} 
		\left( 
    \zeta^2 
    + 1
    \right)^{1/2}
   	\pm \frac{m_2 - m_1}{2} 
   	    \frac{1-|A|^2}{\sqrt{(1-|A|^2)^2 + 4 {\rm Re} A^2}}
   	\right]
   	& \mbox{for the IH case} 
   	\end{array}
   	\right.
\,,
\end{align}
We show in Fig.~\ref{FIG_THsq_OMpm} the maximal and minimum values of $|\Theta_1|^2$ 
by varying the value of $\eta$ as a free parameter.
Note here that $|\Theta_1|^2$ in the considering case
is bounded from below~\cite{Asaka:2015eda} 
by considering $X_\omega =1$ and $\omega_r=0$ as 
\begin{align}
	|\Theta_1|^2 \ge \frac{m_{\ast}}{M_1} \,,\label{Eq:SMALLEST}
\end{align}
where $m_{\ast}=m_2$ or $m_1$ for the NH or IH case, respectively.
This bound is also shown in Fig.~~\ref{FIG_THsq_OMpm} as the black line.
It is thus found that $|\Theta_1|^2$ becomes proportional to 
$M_1$ for $M_1 \gtrsim \Lambda_\beta$, and hence 
a wide region of our possibility can be 
tested by future experiments together with the null observation
of the $0 \nu \beta \beta$ decay.
\begin{figure}[t]
  \centerline{
  \includegraphics[width=7cm]{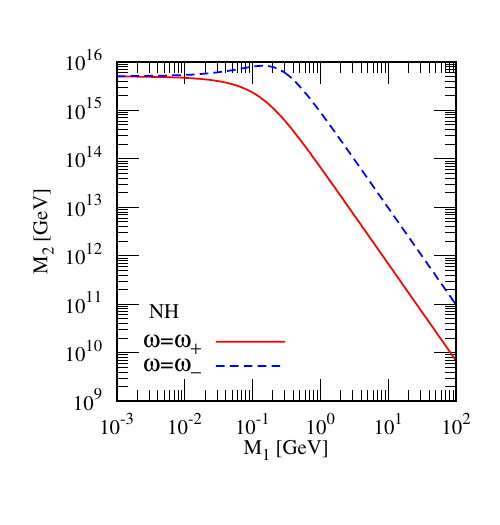}
  \includegraphics[width=7cm]{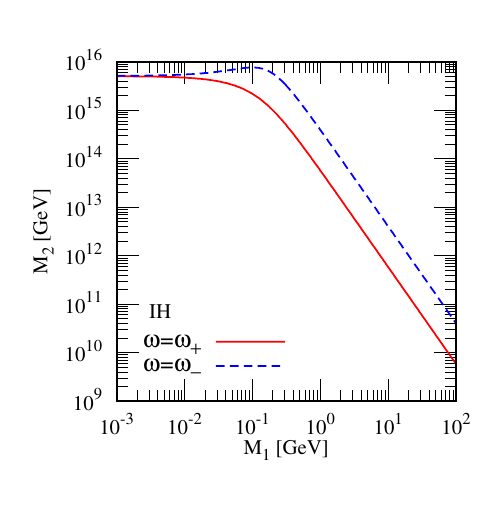}
  }
  \vspace{-2ex}
  \caption{
  	The upper bound on the mass of heavier HNL $N_2$ from
  	the perturbative limit $\sum_\alpha |F_{\alpha 2}|^2 < 4 \pi$ for
  	the NH (left) or IH (right) case.
  	The red solid and blue dashed lines are the bounds
  	for $\omega = \omega_+$ and $\omega_-$, respectively.
  	Here we take $\eta = 0.3 \pi$.
  }
  \label{FIG_MN2_Pert}
\end{figure}

Finally, we mention the properties of heavier HNL $N_2$.
We have assumed so far that its mass is much heavier than $\Lambda_\beta$
so that $N_2$ decouples from the $0 \nu \beta \beta$ decay process. 
On the other hand, since $X_\omega \gg 1$ or $X_\omega^{-1} \gg 1$ as $M_1$ gets heavier,
the Yukawa coupling constants of $N_2$ become rather large 
and exceed the perturbative values when the mass of $N_2$ becomes large. 
See, for example, Ref.~\cite{Asaka:2015eda}. 
Since all other parameters than $M_2$ are already fixed by the conditions related to $N_1$ 
or observables of the neutrino oscillation experiments, 
there is no way to suppress the Yukawa coupling constants of $N_2$. 
\begin{align}
    \sum_\alpha |F_{\alpha 2}|^2
    =
	\left\{
	\begin{array}{ll}
		\displaystyle
    \frac{M_2}{\langle \Phi \rangle^2}
    \left[ 
    \frac{m_3+m_2}{4} \left(X_\omega^2 + X_\omega^{-2} \right)
    + \frac{m_3-m_2}{2} \cos (2 \omega_r)
    \right]	  &
	  \mbox{for the NH case} 
	  \\[3ex]
	  \displaystyle
    \frac{M_2}{\langle \Phi \rangle^2}
    \left[ 
    \frac{m_2+m_1}{4} \left(X_\omega^2 + X_\omega^{-2} \right)
    + \frac{m_2-m_1}{2} \cos (2 \omega_r)
    \right]	  &
	  \mbox{for the IH case} 
	\end{array}
	\right.\,.
\end{align}
Note again that the cancellation in $m_{\rm eff}$
fixes the values of $X_\omega$ and $\omega_r$.
Then, by requiring $\sum_\alpha |F_{\alpha 2}|^2 < 4 \pi$
the upper bound on $M_2$ is obtained which is shown in 
Fig.~\ref{FIG_MN2_Pert}. 
It is seen that the bound becomes stringent for $M_1 \gg \Lambda_\beta$
and $M_2 < {\cal O}(10^{10} \mathchar`- 10^{11})$~GeV when $M_1 = 10^2$~GeV 
simply because it makes $X_\omega$ to be the largest in the considered mass range of $M_1$.

\section{Discussions and conclusions}
\label{sec:conclusions}

In this paper we have examined the neutrinoless double beta decay in 
the seesaw mechanism with two right-handed neutrinos.
The Majorana nature of active neutrinos and heavy neutral leptons breaks the lepton number of the theory,
which may lead to the neutrinoless double beta decay. 
In this framework, the cancellation among active neutrino contributions 
cannot occur since the lightest active neutrino is massless. 
heavy neutral leptons can give a significant contribution in addition to those from active neutrinos, 
and the effective neutrino mass can vanish in some cases.  
One possibility is that all the heavy neutral leptons are lighter than $\Lambda_\beta$.
It has been shown that the lighter heavy neutral lepton $N_1$ can obliterate 
the neutrinoless double beta decay even if active neutrinos do contribute it.
In such a case the required mixing elements becomes proportional to the mass
for $M_1 \gtrsim 0.1$~GeV. 
Then, the future experiments like
DUNE, SHiP, and FCC-ee can test the present scenario even if 
the neutrinoless double beta decay processes are not observed. 

If $N_1$ becomes light enough, the suggested values of the lifetime are so long that 
$N_1$ decays after the onset of the big-bang nucleosynthesis (BBN) 
and would destroy the success of the BBN and/or conflict with the observational
data of the cosmic microwave background radiation. 
One possibility to avoid this difficulty is the dilution of the $N_1$ abundance
by the late time entropy production.   Such an additional production may be
realized by the decay of the heavier heavy neutral lepton $N_2$~\cite{Asaka:2006ek}. 
Such a bound, however, becomes safe enough when $M_1 \gtrsim 1~{\rm GeV}$. 
Namely, the suppression in the neutrinoless double beta decay rate 
can be easily achieved in wide region of $M_1$ 
without conflicting any constraints at the moment. 

It has also been shown that 
even if the neutrinoless double beta decay at an experiment, 
namely $N_1$ contribution conceals the decay in a specific element, 
there is still a possibility to detect the decay at other experiments which use other nuclei. 
If the neutrinoless double beta decay have been observed at an experiment while another one cannot do, 
it may mean the missing decay signal is caused by the heavy neutral lepton contribution.

\section*{Acknowledgments}
The authors thank Dr. Shintaro Eijima for the discussion on the topic 
in the Appendix~\ref{sec:ap1}. 
The work of T.A. was partially supported by JSPS KAKENHI
Grant Numbers 17K05410, 18H03708, 19H05097, and 20H01898.
The work of H.I. was supported by JSPS KAKENHI Grant Numbers 18H03708.

\appendix
\section{Physical region of the model parameters}
\label{sec:ap1}

In this Appendix we explain the physical region of the model parameters,
especially, parameters for the right-handed neutrinos based on the
discussion in Ref.~\cite{deGouvea:2000pqg}.  First of all, we observe
that the Lagrangian is invariant under $\omega \to \omega + \pi$,
$N_1 \to - N_1$ and $N_2 \to - N_2$. 
Thus, the two ranges of
$\rew$, $\left[ - \frac{\pi}{2}, \frac{\pi}{2} \right]$ and
$\left[ \frac{\pi}{2}, \frac{3 \pi}{2} \right]$, are physically
equivalent and it is enough to consider the range
$\left[ - \frac{\pi}{2}, \frac{\pi}{2} \right]$.  Second, the
Lagrangian is also invariant under $\omega \to \frac \pi 2 - \omega$,
$\xi \to - \xi$, 
$\nu_{R1} \to - \nu_{R2}$ and $\nu_{R2} \to - \nu_{R1}$ 
(\ie, the change of
the mass ordering $M_2 > M_1$ $\to$ $M_1 < M_2$).  It is therefore
found that we can take $M_2 > M_1$ without loss of generality.  Third,
the change of Majorana phase $\eta \to \eta + \pi$ can be compensated by
$\xi \to - \xi$, and then the physical range of $\eta$ 
is taken to be $[0, \pi]$ in general.
Finally, 
a set of the transformations $\omega \to - \omega$, $\xi \to - \xi$ and
$\nu_{R2} \to - \nu_{R2}$ is another symmetry of the Lagrangian,
which restricts the physical range of the parameter space.

We find that there are 16 independent choices
for the physical regions.  Here we show the three 
useful choices:
\begin{itemize}
\item[(1)] Fix $\xi = + 1$ (or $\xi = -1$), consider
the both ranges of $\rew$  $\left[ - \frac{\pi}{2}, 0 \right]$
and $\left[ 0, \frac{\pi}{2} \right]$, and 
consider the whole range of $\imw$ 
$\left[ - \infty, \infty \right]$.
\item[(2)] Consider both signs of $\xi$,
consider the half range of $\rew$ $\left[ 0, \frac{\pi}{2} \right]$
(or $\left[ - \frac{\pi}{2}, 0 \right]$),
and the whole range of $\imw$ 
$\left[ - \infty, \infty \right]$.
\item[(3)] Consider both signs of $\xi$,
consider the both ranges of $\rew$ $\left[ 0, \frac{\pi}{2} \right]$
and $\left[ - \frac{\pi}{2}, 0 \right]$,
and the half range of $\imw$ 
$\left[ 0, \infty \right]$ (or $\left[ -\infty, 0 \right]$).
\end{itemize}
In the analysis~\cite{Asaka:2011pb} 
the option (3) has been selected.
These choices of the physical region of parameters
$\xi$, $\omega$ and $\eta$ are summarized in Tab.~\ref{tab:physical_region}.
In the present analysis we choose the option (1) by fixing 
the sign parameter $\xi = + 1$.
\begin{table}[h]
  \begin{center}
$\renewcommand{\arraystretch}{1.8}
\begin{array}{ | c || c | c | c | c || c |}
   \hline
   {} & \xi & \omega_r & \omega_i & X_\omega = \exp (\omega_i) & \eta 
   \\ \hline   
   \mbox{(1)} & +1 & \left[ - \frac \pi 2, \frac \pi 2 \right] & \left[ - \infty, + \infty \right] & [ 0 , + \infty] & [ 0, \pi ] 
   \\ \hline   
   \mbox{(2)} & \pm 1 & \left[ 0, \frac \pi 2 \right] & \left[ - \infty, + \infty \right] & [ 0 , + \infty] & [ 0, \pi ] 
   \\ \hline   
   \mbox{(3)} & \pm 1 & \left[ - \frac \pi 2, \frac \pi 2 \right] & \left[ 0, + \infty \right] & [ 1 , + \infty] & [ 0, \pi ] 
   \\ \hline   
    \end{array}
    $
  \end{center}
    \caption{
      Three options of the physical region of 
      parameters $\xi$, $\rew$, $\imw$, 
      $X_\omega$ and $\eta$.
      }
    \label{tab:physical_region}
\end{table}


\section{Upper and lower bounds of mixing elements in each flavor}
\label{sec:ap3}
\begin{figure}[ht]
  \centerline{
  \includegraphics[width=6cm]{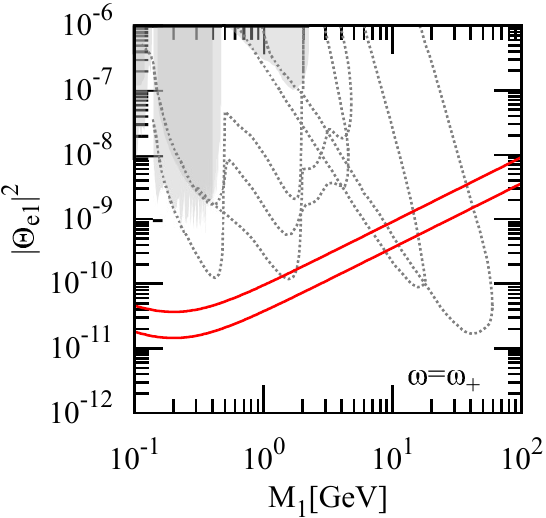}%
  \includegraphics[width=6cm]{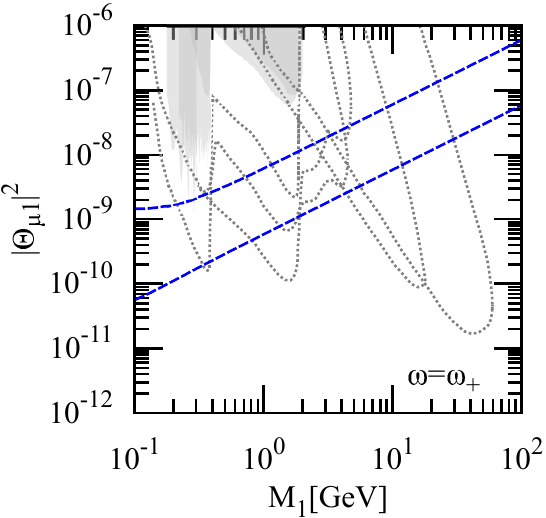}%
  \includegraphics[width=6cm]{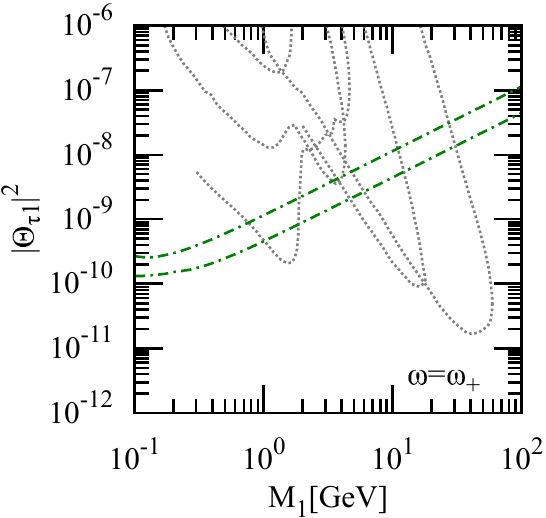}%
  }%
  \centerline{
  \includegraphics[width=6cm]{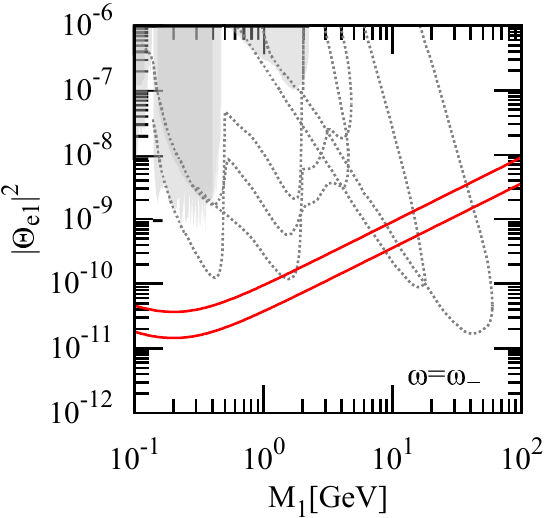}%
  \includegraphics[width=6cm]{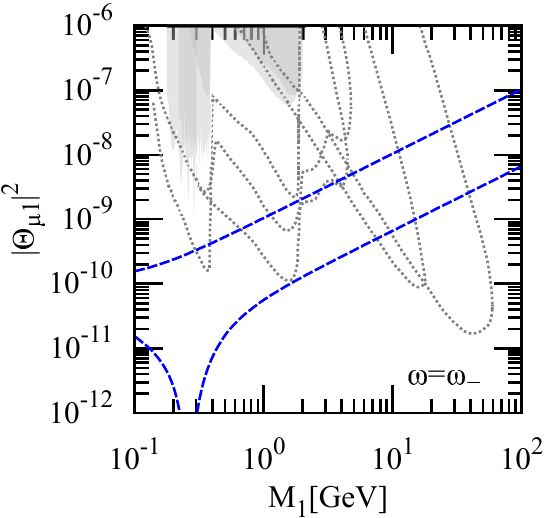}%
  \includegraphics[width=6cm]{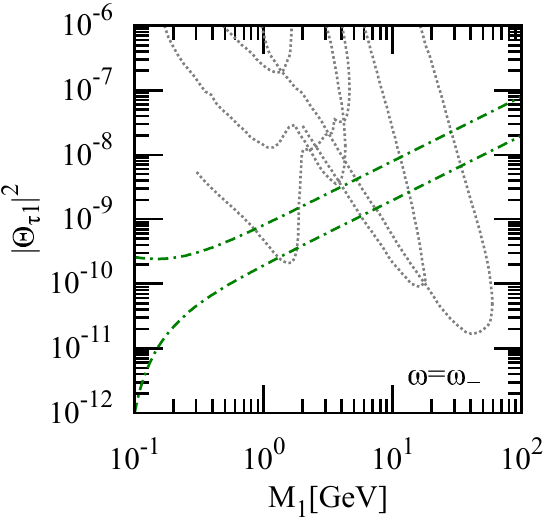}%
  }%

  \vspace{-2ex}
  \caption{
  {Upper and lower bounds on 
  the mixing elements $|\Theta_{e1}|^2$ (left, red solid lines),
  $|\Theta_{\mu 1}|^2$ (middle, blue dashed lines) and
  $|\Theta_{\tau 1}|^2$ (right, green dot-dashed lines) 
  for vanishing $m_{\rm eff}$ in the NH case.
  Here take $\omega = \omega_+$ (upper panel) and 
  $\omega_-$ (lower panel).}
  }
  \label{fig:FIG_THSQ_NH}
\end{figure}

\begin{figure}[ht]
  \centerline{
  \includegraphics[width=6cm]{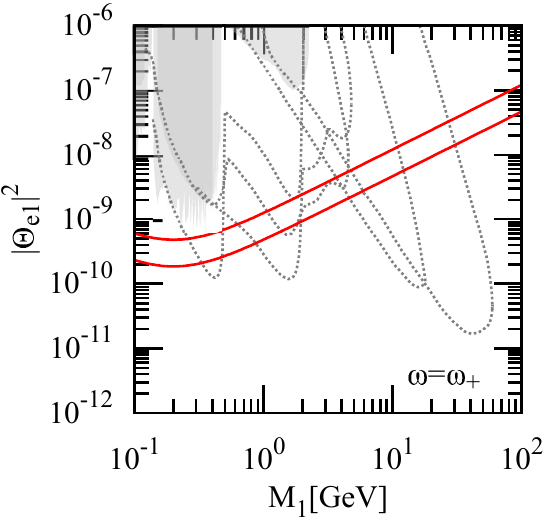}%
  \includegraphics[width=6cm]{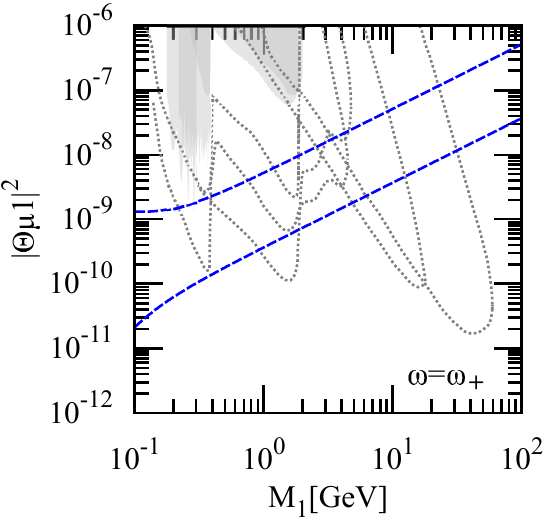}%
  \includegraphics[width=6cm]{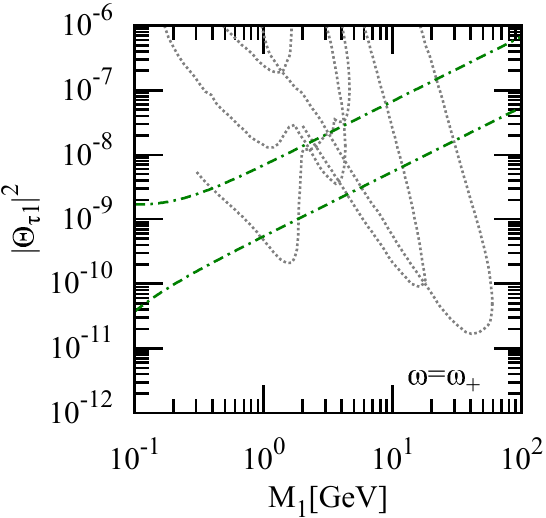}%
  }%
  \centerline{
  \includegraphics[width=6cm]{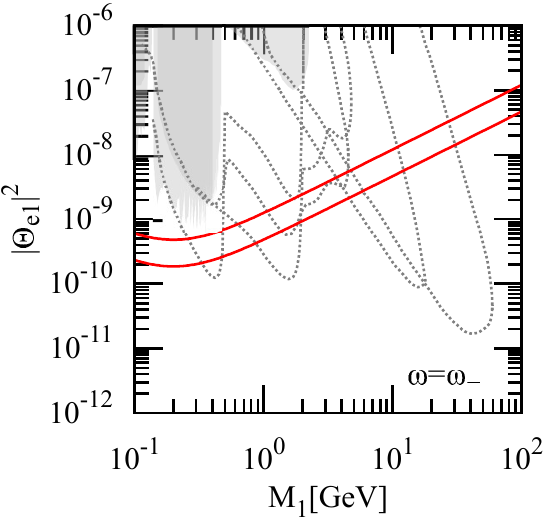}%
  \includegraphics[width=6cm]{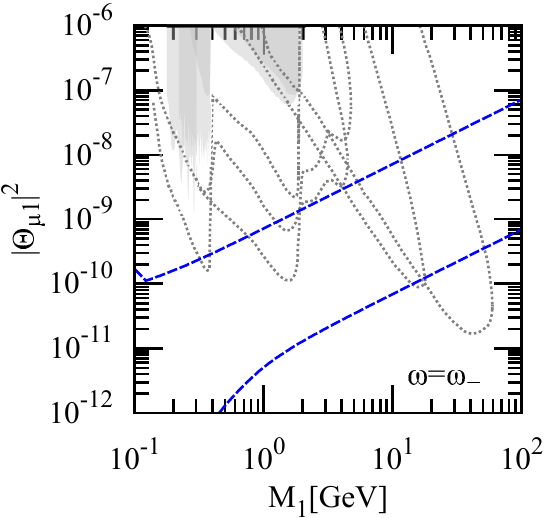}%
  \includegraphics[width=6cm]{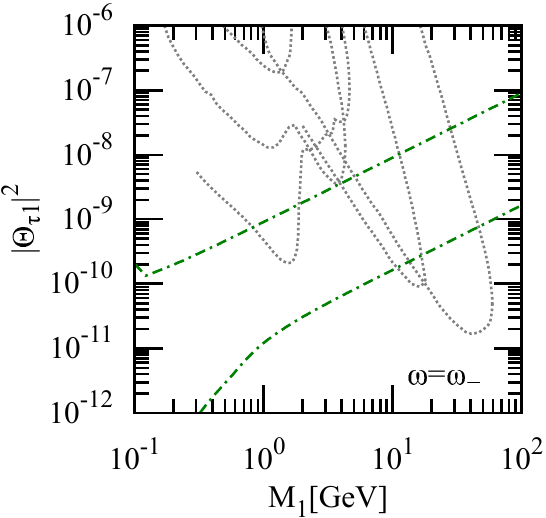}%
  }%

  \vspace{-2ex}
  \caption{
  {Upper and lower bounds on 
  the mixing elements $|\Theta_{e1}|^2$ (left, red solid lines),
  $|\Theta_{\mu 1}|^2$ (middle, blue dashed lines) and
  $|\Theta_{\tau 1}|^2$ (right, green dot-dashed lines) 
  for vanishing $m_{\rm eff}$ in the IH case.
  Here take $\omega = \omega_+$ (upper panel) and 
  $\omega_-$ (lower panel).}
  }
  \label{fig:FIG_THSQ_IH}
\end{figure}

As shown in the main text, the cancellation condition 
is directly tied to the value of the electron-type mixing element, 
whereas mixing elements of other flavor can also be determined 
since all the free parameters such as $\omega$ can be fixed. 
In this Appendix, we show the predicted 
values of mixing elements for all flavor components 
in Figs.~\ref{fig:FIG_THSQ_NH} and \ref{fig:FIG_THSQ_IH} for the NH and IH case, respectively, 
together with the regions where have already been excluded by previous experiments 
and the future experiments can search. 

\clearpage


\begin{thebibliography}{100}

\bibitem{Minkowski:1977sc} 
  P.~Minkowski,
  Phys.\ Lett.\ B {\bf 67}, 421 (1977).

\bibitem{Yanagida:1979as} 
T. Yanagida, in Proceedings of the Workshop on Unified Theory and Baryon Number of the Universe, edited by.O. Sawada and A. Sugamoto (KEK, Tsukuba, Ibaraki 305- 0801 Japan, 1979) p. 95.

\bibitem{Yanagida:1980xy} 
  T.~Yanagida,
  Prog.\ Theor.\ Phys.\  {\bf 64}, 1103 (1980).

%
\bibitem{Ramond:1979py}
P.~Ramond,
[arXiv:hep-ph/9809459 [hep-ph]].

\bibitem{GellMann:1980vs} 
  M. Gell-Mann, P. Ramond, and R. Slansky, in Supergravity, edited by.P. van Niewwenhuizen and D. Freedman (North Holland, Amsterdam, 1979)
  [arXiv:1306.4669 [hep-th]].

\bibitem{Glashow:1979nm}
S.~L.~Glashow,
NATO Sci. Ser. B \textbf{61} (1980), 687
doi:10.1007/978-1-4684-7197-7\_15

\bibitem{Mohapatra:1979ia}
  R.~N.~Mohapatra and G.~Senjanovic,
  Phys.\ Rev.\ Lett.\  {\bf 44} (1980) 912.


\bibitem{tHooft:1976rip}
G.~'t Hooft,
Phys. Rev. Lett. \textbf{37} (1976), 8-11.


\bibitem{tHooft:1976snw}
G.~'t Hooft,
Phys. Rev. D \textbf{14} (1976), 3432-3450
doi:10.1103/PhysRevD.14.3432

\bibitem{Doi:1985dx}
M.~Doi, T.~Kotani and E.~Takasugi,
Prog. Theor. Phys. Suppl. \textbf{83} (1985), 1
doi:10.1143/PTPS.83.1

\bibitem{Pas:2015eia}
H.~P\"as and W.~Rodejohann,
New J. Phys. \textbf{17} (2015) no.11, 115010
doi:10.1088/1367-2630/17/11/115010
[arXiv:1507.00170 [hep-ph]].

\bibitem{DellOro:2016tmg}
S.~Dell'Oro, S.~Marcocci, M.~Viel and F.~Vissani,
Adv. High Energy Phys. \textbf{2016} (2016), 2162659
doi:10.1155/2016/2162659
[arXiv:1601.07512 [hep-ph]].

\bibitem{Dolinski:2019nrj}
M.~J.~Dolinski, A.~W.~Poon and W.~Rodejohann,
Ann. Rev. Nucl. Part. Sci. \textbf{69} (2019), 219-251
doi:10.1146/annurev-nucl-101918-023407
[arXiv:1902.04097 [nucl-ex]].

\bibitem{Rizzo:1982kn}
T.~G.~Rizzo,
Phys. Lett. B \textbf{116} (1982), 23-28
doi:10.1016/0370-2693(82)90027-2

\bibitem{London:1987nz}
D.~London, G.~Belanger and J.~N.~Ng,
Phys. Lett. B \textbf{188} (1987), 155-158
doi:10.1016/0370-2693(87)90723-4.

\bibitem{Dicus:1991fk}
D.~A.~Dicus, D.~D.~Karatas and P.~Roy,
Phys. Rev. D \textbf{44} (1991), 2033-2037
doi:10.1103/PhysRevD.44.2033.

\bibitem{Belanger:1995nh}
G.~Belanger, F.~Boudjema, D.~London and H.~Nadeau,
Phys. Rev. D \textbf{53} (1996), 6292-6301
doi:10.1103/PhysRevD.53.6292
[arXiv:hep-ph/9508317 [hep-ph]].

\bibitem{Gluza:1995ky}
J.~Gluza and M.~Zralek,
Phys. Rev. D \textbf{52} (1995), 6238-6248
doi:10.1103/PhysRevD.52.6238
[arXiv:hep-ph/9502284 [hep-ph]].

\bibitem{Gluza:1995ix}
J.~Gluza and M.~Zralek,
Phys. Lett. B \textbf{362} (1995), 148-154
doi:10.1016/0370-2693(95)01158-M
[arXiv:hep-ph/9507269 [hep-ph]].


\bibitem{Gluza:1995js}
J.~Gluza and M.~Zralek,
Phys. Lett. B \textbf{372} (1996), 259-264
doi:10.1016/0370-2693(96)00074-3
[arXiv:hep-ph/9510407 [hep-ph]].

\bibitem{Greub:1996ct}
C.~Greub and P.~Minkowski,
eConf \textbf{C960625} (1996), NEW149
doi:10.1142/S0217751X98001153
[arXiv:hep-ph/9612340 [hep-ph]].

\bibitem{Rodejohann:2010jh}
W.~Rodejohann,
Phys. Rev. D \textbf{81} (2010), 114001
doi:10.1103/PhysRevD.81.114001
[arXiv:1005.2854 [hep-ph]].

\bibitem{Banerjee:2015gca}
S.~Banerjee, P.~S.~B.~Dev, A.~Ibarra, T.~Mandal and M.~Mitra,
Phys. Rev. D \textbf{92} (2015), 075002
doi:10.1103/PhysRevD.92.075002
[arXiv:1503.05491 [hep-ph]].


\bibitem{Asaka:2015oia}
T.~Asaka and T.~Tsuyuki,
Phys. Rev. D \textbf{92} (2015) no.9, 094012
doi:10.1103/PhysRevD.92.094012
[arXiv:1508.04937 [hep-ph]].


\bibitem{Wang:2016eln}
K.~Wang, T.~Xu and L.~Zhang,
Phys. Rev. D \textbf{95} (2017) no.7, 075021
doi:10.1103/PhysRevD.95.075021
[arXiv:1610.02618 [hep-ph]].

\bibitem{Ilakovac:1995km}
A.~Ilakovac, B.~A.~Kniehl and A.~Pilaftsis,
Phys. Rev. D \textbf{52} (1995), 3993-4005
doi:10.1103/PhysRevD.52.3993
[arXiv:hep-ph/9503456 [hep-ph]].
\bibitem{Ilakovac:1994kj}
A.~Ilakovac and A.~Pilaftsis,
Nucl. Phys. B \textbf{437} (1995), 491
doi:10.1016/0550-3213(94)00567-X
[arXiv:hep-ph/9403398 [hep-ph]].
\bibitem{Ilakovac:1995wc}
A.~Ilakovac,
Phys. Rev. D \textbf{54} (1996), 5653-5673
doi:10.1103/PhysRevD.54.5653
[arXiv:hep-ph/9608218 [hep-ph]].
\bibitem{Gribanov:2001vv}
V.~Gribanov, S.~Kovalenko and I.~Schmidt,
Nucl. Phys. B \textbf{607} (2001), 355-368
doi:10.1016/S0550-3213(01)00169-9
[arXiv:hep-ph/0102155 [hep-ph]].
\bibitem{Atre:2005eb}
A.~Atre, V.~Barger and T.~Han,
Phys. Rev. D \textbf{71} (2005), 113014
doi:10.1103/PhysRevD.71.113014
[arXiv:hep-ph/0502163 [hep-ph]].



\bibitem{Ng:1978ij}
J.~N.~Ng and A.~N.~Kamal,
Phys. Rev. D \textbf{18} (1978), 3412
doi:10.1103/PhysRevD.18.3412.
\bibitem{Abad:1984gh}
J.~Abad, J.~G.~Esteve and A.~F.~Pacheco,
Phys. Rev. D \textbf{30} (1984), 1488
doi:10.1103/PhysRevD.30.1488.
\bibitem{Dib:2000wm}
C.~Dib, V.~Gribanov, S.~Kovalenko and I.~Schmidt,
Phys. Lett. B \textbf{493} (2000), 82-87
doi:10.1016/S0370-2693(00)01134-5
[arXiv:hep-ph/0006277 [hep-ph]].
\bibitem{Ali:2001gsa}
A.~Ali, A.~V.~Borisov and N.~B.~Zamorin,
Eur. Phys. J. C \textbf{21} (2001), 123-132
doi:10.1007/s100520100702
[arXiv:hep-ph/0104123 [hep-ph]].
\bibitem{Asaka:2016rwd}
T.~Asaka and H.~Ishida,
Phys. Lett. B \textbf{763} (2016), 393-396
doi:10.1016/j.physletb.2016.10.070
[arXiv:1609.06113 [hep-ph]].

\bibitem{KamLAND-Zen:2016pfg}
A.~Gando \textit{et al.} [KamLAND-Zen],
Phys. Rev. Lett. \textbf{117} (2016) no.8, 082503
doi:10.1103/PhysRevLett.117.082503
[arXiv:1605.02889 [hep-ex]].

%
\bibitem{Blennow:2010th}
M.~Blennow, E.~Fernandez-Martinez, J.~Lopez-Pavon and J.~Menendez,
JHEP \textbf{07} (2010), 096
doi:10.1007/JHEP07(2010)096
[arXiv:1005.3240 [hep-ph]].


\bibitem{Asaka:2020wfo}
T.~Asaka, H.~Ishida and K.~Tanaka,
[arXiv:2012.12564 [hep-ph]].

\bibitem{Pontecorvo:1957qd}
B.~Pontecorvo,
Sov. Phys. JETP \textbf{7} (1958), 172-173

\bibitem{Maki:1962mu}
  Z.~Maki, M.~Nakagawa and S.~Sakata,
  Prog.\ Theor.\ Phys.\  {\bf 28} (1962) 870.


\bibitem{Casas:2001sr}
  J.~A.~Casas and A.~Ibarra,
  Nucl.\ Phys.\  B {\bf 618} (2001) 171
  [arXiv:hep-ph/0103065].

\bibitem{Abada:2006ea} 
  A.~Abada, S.~Davidson, A.~Ibarra,
  F.~X.~Josse-Michaux, M.~Losada and A.~Riotto,
  JHEP {\bf 0609} (2006) 010
  [arXiv:hep-ph/0605281].

\bibitem{Asaka:2011pb}
  T.~Asaka, S.~Eijima and H.~Ishida,
  JHEP {\bf 1104} (2011) 011
  doi:10.1007/JHEP04(2011)011
  [arXiv:1101.1382 [hep-ph]].

%
\bibitem{nufit}
I.~Esteban, M.~Gonzalez-Garcia, A.~Hernandez-Cabezudo, M.~Maltoni and T.~Schwetz,
JHEP \textbf{01} (2019), 106
doi:10.1007/JHEP01(2019)106
[arXiv:1811.05487 [hep-ph]],
``NuFiT 5.0: Three-neutrino fit based on data available in July 2020,''
www.nu-fit.org.

\bibitem{Faessler:2014kka}
A.~Faessler, M.~Gonzlez, S.~Kovalenko and F.~Simkovic,
Phys. Rev. D \textbf{90} (2014) no.9, 096010
doi:10.1103/PhysRevD.90.096010
[arXiv:1408.6077 [hep-ph]].

\bibitem{Barea:2015zfa}
J.~Barea, J.~Kotila and F.~Iachello,
Phys. Rev. D \textbf{92} (2015), 093001
doi:10.1103/PhysRevD.92.093001
[arXiv:1509.01925 [hep-ph]].

%


%
\bibitem{Halprin:1983ez}
A.~Halprin, S.~T.~Petcov and S.~P.~Rosen,
Phys. Lett. B \textbf{125} (1983), 335-338
doi:10.1016/0370-2693(83)91296-0

%
\bibitem{Leung:1984vy}
C.~N.~Leung and S.~T.~Petcov,
Phys. Lett. B \textbf{145} (1984), 416-420
doi:10.1016/0370-2693(84)90071-6

%
\bibitem{PIENU:2011aa}
M.~Aoki \textit{et al.} [PIENU],
Phys. Rev. D \textbf{84} (2011), 052002
doi:10.1103/PhysRevD.84.052002
[arXiv:1106.4055 [hep-ex]].

%
\bibitem{Aguilar-Arevalo:2017vlf}
A.~Aguilar-Arevalo \textit{et al.} [PIENU],
Phys. Rev. D \textbf{97} (2018) no.7, 072012
doi:10.1103/PhysRevD.97.072012
[arXiv:1712.03275 [hep-ex]].


%
\bibitem{NA62:2020mcv}
E.~Cortina Gil \textit{et al.} [NA62],
Phys. Lett. B \textbf{807} (2020), 135599
doi:10.1016/j.physletb.2020.135599
[arXiv:2005.09575 [hep-ex]].


%
\bibitem{Blondel:2014bra}
A.~Blondel \textit{et al.} [FCC-ee study Team],
Nucl. Part. Phys. Proc. \textbf{273-275} (2016), 1883-1890
doi:10.1016/j.nuclphysbps.2015.09.304
[arXiv:1411.5230 [hep-ex]].

%
\bibitem{SHiP:2018xqw}
C.~Ahdida \textit{et al.} [SHiP],
JHEP \textbf{04} (2019), 077
doi:10.1007/JHEP04(2019)077
[arXiv:1811.00930 [hep-ph]].

%
\bibitem{Krasnov:2019kdc}
I.~Krasnov,
Phys. Rev. D \textbf{100} (2019) no.7, 075023
doi:10.1103/PhysRevD.100.075023
[arXiv:1902.06099 [hep-ph]].

%
\bibitem{Alpigiani:2020tva}
C.~Alpigiani \textit{et al.} [MATHUSLA],
[arXiv:2009.01693 [physics.ins-det]].

\bibitem{Asaka:2015eda}
T.~Asaka and T.~Tsuyuki,
Phys. Lett. B \textbf{753} (2016), 147-149
doi:10.1016/j.physletb.2015.12.013
[arXiv:1509.02678 [hep-ph]].

\bibitem{Asaka:2006ek}
T.~Asaka, M.~Shaposhnikov and A.~Kusenko,
Phys. Lett. B \textbf{638} (2006), 401-406
doi:10.1016/j.physletb.2006.05.067
[arXiv:hep-ph/0602150 [hep-ph]].


\bibitem{deGouvea:2000pqg}
  A.~de Gouvea, A.~Friedland and H.~Murayama,
  Phys.\ Lett.\ B {\bf 490} (2000) 125
  [hep-ph/0002064].



\end{thebibliography}
\end{document}